\documentclass{amsip-p-l}
\usepackage{mathrsfs}
\usepackage{amssymb}
\usepackage{amsfonts}

\theoremstyle{definition}

\theoremstyle{remark}

\numberwithin{equation}{section}

\def\Fg{{\mathfrak g}}
\def\Fh{{\mathfrak h}}
\def\Ft{{\mathfrak t}}
\def\FL{{\mathfrak L}}
\def\BC{{\mathbb C}}
\def\BP{{\mathbb P}}
\def\BR{{\mathbb R}}
\def\BZ{{\mathbb Z}}
\def\RC{{\mathrm C}}
\def\RS{{\mathrm S}}
\def\CA{{\mathcal A}}
\def\CD{{\mathcal D}}
\def\CF{{\mathcal F}}
\def\CG{{\mathcal G}}

\def\CO{{\mathcal O}}
\def\CS{{\mathcal S}}
\def\CW{{\mathcal W}}
\def\Rc{{\mathrm c}}
\def\Rs{{\mathrm s}}
\def\RZ{{\mathrm Z}}
\def\RR{{\mathrm R}}

\def\SH{{\mathscr H}}

\def\^{{\wedge}}
\def\*{{\star}}
\def\Hol{{\mathop{\rm Hol}}}
\def\Tr{{\mathop{\rm Tr}}}
\def\Lie{{\mathop{\rm Lie}}}
\def\Vol{{\mathop{\rm Vol}}}
\def\ad{{\mathop{\rm ad}}}
\def\ch{{\mathop{\rm ch}}}
\def\e#1{{\rm e}^{\, #1}}

\def\ha{{1\over 2}}

\def\bar{\overline}
\def\ni{\noindent}

\begin{document}

\title{Remarks on Wilson Loops and Seifert Loops in Chern-Simons Theory}

\author{Chris Beasley}

\address{Simons Center for Geometry and Physics, Stony Brook
  University, Stony Brook, New York 11794-3636}

\email{cbeasley@insti.physics.sunysb.edu}


\begin{abstract}
As noted long ago by Atiyah and Bott, the classical Yang-Mills action
on a Riemann surface admits a beautiful symplectic interpretation as
the norm-square of a moment map associated to the Hamiltonian action
by gauge transformations on the affine space of connections.  Here I
will explain how certain Wilson loop observables in Chern-Simons gauge
theory on a Seifert three-manifold can be given an analogous
symplectic description.  Among other results, this symplectic
description implies that the stationary-phase approximation to the
Wilson loop path integral is exact for torus knots, an empirical observation
made previously by Lawrence and Rozansky.  This article reviews
selected material from the larger work ``Localization for Wilson Loops
in Chern-Simons Theory,'' {\ttfamily arXiv:0911.2687}.
\end{abstract}

\maketitle

\section{Introduction}

This brief article is intended as an introduction to the study of
Chern-Simons gauge theory via non-abelian localization
\cite{Beasley:2005,Beasley:2009}.  I will not attempt to give here
a comprehensive overview of the subject.  Instead, my goal is to
highlight two very beautiful ideas, one old and one new, which enter
the story of non-abelian localization in an essential way and which
may have applications elsewhere.  See also
\cite{Blau:2006gh,Kapustin:2009kz} for alternative approaches to path
integral localization in Chern-Simons theory. 

To set the stage, we consider Chern-Simons theory on a compact,
oriented three-manifold $M$ with gauge group $G$.  By assumption $G$
will be a compact, connected, simply-connected, and simple Lie group.  For
instance, $G$ could be $SU(N)$ for any ${N > 1}$.  These assumptions
on $G$ ensure that any principal $G$-bundle $P$ over $M$ is
topologically trivial, a technical convenience.  The gauge field $A$
of Chern-Simons theory is then a connection on $P$.

Let me now introduce the Wilson loop operators in Chern-Simons theory.
Quite generally, a Wilson loop operator $W_R(C)$ in any gauge theory
on a manifold $M$ is described by the data of an oriented, closed
curve $C$ which is smoothly\footnote{The condition that $C$ be
  smoothly embedded in $M$ is merely for convenience and is not 
  strictly required to define $W_R(C)$ as a sensible operator in gauge  
  theory.  Indeed, the Wilson loop expectation value in Chern-Simons
  theory can be computed exactly even for the case that $C$ is an
  arbitrary closed graph \cite{WittenWF:1989} in $M$.} 
embedded in $M$ and which is decorated by an irreducible
representation $R$ of the gauge group $G$.  As a classical functional of
the connection $A$, the Wilson loop operator $W_R(C)$ is then given simply 
by the trace in $R$ of the holonomy of $A$ around $C$,
\begin{equation}\label{WRC}
\begin{split}
W_R(C) \,&=\, \Tr_R \Hol_C(A)\,,\\
&=\, \Tr_R P\exp{\!\left(-\oint_C A\right)}\,. 
\end{split}
\end{equation}
Following the standard practice in physics, we describe the holonomy
$\Hol_C(A)$ in the second line of \eqref{WRC} in terms of a 
path-ordered exponential $P\exp{\!(-\oint_C A)}$, which describes solutions 
to the first-order differential equation for parallel transport\footnote{Because we work in conventions for which ${d_A = d + A}$ is the covariant derivative, a minus sign appears in the argument of the path-ordered exponential.} along $C$.

With the Wilson loop operator in hand, we finally introduce the
absolutely-normalized Wilson loop path integral  
\begin{equation}\label{ZkCR}
Z(k; C, R) =\, \frac{1}{\Vol(\CG)} \int_\CA {\mathcal D}\!A \,\,
W_R(C) \, \exp{\!\left[i \, \frac{k}{4\pi} \, \RC\RS(A)\right]}\,.
\end{equation}
Here $\RC\RS(A)$ is the Chern-Simons action,
\begin{equation}
\RC\RS(A) = \int_M \Tr\!\left( A\^dA \,+\, \frac{2}{3} A \^ A \^ A\right),
\end{equation}
and `$\Tr$' denotes a suitably-normalized, negative-definite,
invariant quadratic form on the Lie algebra $\Fg$ of $G$.  For
instance, if $G$ is $SU(N)$, the quadratic form `$\Tr$' is given
concretely by the trace in the fundamental $N$-dimensional
representation of $SU(N)$.

For later use, let me recall three essential properties of the
Chern-Simons action.  First, the Chern-Simons action is purely
topological, insofar as it depends only on the choice of an
orientation, not a metric, on $M$.  Second, the critical points of the
Chern-Simons action are precisely the flat connections on $M$, for
which 
\begin{equation}
F_A = dA + A\^A = 0.
\end{equation}
Finally, even though the Chern-Simons action is not manifestly
gauge-invariant ---and indeed the Chern-Simons action is {\em not}
gauge-invariant --- the Chern-Simons action is almost gauge-invariant,
in the sense that it is invariant under homotopically trivial gauge
transformations.  Under homotopically nontrivial gauge
transformations, the functional $\RC\RS(A)$ shifts by integral
multiples of $8\pi^2$, where the relevant integer is determined by the
homotopy class of the gauge transformation as a map from $M$ to $G$.
So long as the Chern-Simons level ${k\in\BZ}$ is quantized to be an integer,
the exponential of the Chern-Simons action in \eqref{ZkCR} is then honestly 
gauge-invariant.

Otherwise, we present $Z(k; C, R)$ as an integral over the 
infinite-dimensional affine space $\CA$ of connections on $M$.  As usual in gauge theory, we divide that integral by the volume of the infinite-dimensional group 
$\CG$ of gauge transformations acting on $\CA$.

Before proceeding, let me make one elementary remark.  When $R$ is the 
trivial representation, the Wilson loop operator $W_R(C)$ is the identity operator, and the absolutely-normalized Wilson loop path integral reduces immediately to the path integral which describes the Chern-Simons partition function $Z(k)$ on $M$,
\begin{equation}\label{CSZ}
Z(k) =\, \frac{1}{\Vol(\CG)} \int_\CA {\mathcal D}\!A \,\,
\exp{\!\left[i \, \frac{k}{4\pi} \, \RC\RS(A)\right]}\,.
\end{equation}
In terms of $Z(k)$ and $Z(k; C, R)$, the Wilson loop expectation value 
$\big\langle W_R(C)\big\rangle$ is then given by the ratio 
\begin{equation}
\big\langle W_R(C)\big\rangle = \frac{Z(k;C,R)}{Z(k)}.
\end{equation}
Though the Wilson loop expectation value is very convenient to consider for 
certain purposes, we work exclusively with $Z(k;C,R)$ today.

As it stands, the Wilson loop path integral in \eqref{ZkCR} is a purely formal
expression.  Nonetheless, over twenty years ago Witten
\cite{Witten:1988} gave a completely precise prescription to define
the observable $Z(k; C, R)$, based upon the canonical
quantization of Chern-Simons theory in the Hamiltonian formalism.
This formalism leads to a well-known algebraic description
\cite{Reshetikhin:1991} of $Z(k; C, R)$ in terms of a presentation of
$M$ via surgery on $S^3$, combined with data about certain modular representations associated to two-dimensional rational conformal field theories.

Beyond providing a rigorous means to define the observable $Z(k; C,
R)$, the Hamiltonian formalism is also very powerful, insofar as it
can be used to perform explicit, exact calculations in Chern-Simons
theory.  For instance, among the classic results in
\cite{Witten:1988}, Witten computed the unknot observable  
$Z\big(k; \bigcirc, \mbox{\boldmath$j$}\big)$ for Chern-Simons theory on
$S^3$ with gauge group $SU(2)$, 
\begin{equation}\label{UNKNOTW}
Z\big(k; \bigcirc, \mbox{\boldmath$j$}\big) \,=\, \sqrt{\frac{2}{k+2}} \,
\sin\!\left(\frac{\pi \, j}{k+2}\right),\qquad\qquad j =
1,\ldots,k+1\,.
\end{equation}
Here $\mbox{\boldmath$j$}$ is the irreducible representation of $SU(2)$ with
dimension $j$, and as indicated, $\mbox{\boldmath$j$}$ runs without
loss over the finite set of representations which are integrable in
the affine Lie algebra for $SU(2)$ at level $k$. 

Yet despite its computability, the algebraic definition of $Z(k; C, R)$ in
the Hamiltonian formalism obscures many features which are manifest in
the preceding path integral \eqref{ZkCR} and which one would like to
understand more deeply.  As a simple example, in the semi-classical
limit that $k$ is large, a naive stationary-phase approximation can be
applied to the path integral describing the Chern-Simons partition function,
and this approximation implies asymptotic behavior for $Z(k)$ that is
far from evident in the complicated, exact expressions that arise from
conformal field theory.  Nonetheless, the predicted asymptotic
behavior can be checked in examples, as was done early on by
Freed and Gompf \cite{Freed:1991}, Jeffrey \cite{Jeffrey:1992}, and
Garoufalidis \cite{Garoufalidis:1992}.  See for instance \S 7 of
\cite{Ohtsuki:2001} for a survey of continuing work in this area.

This article concerns a very special and very beautiful situation in which the
stationary-phase approximation to the Wilson loop path integral is
actually exact.  Namely, the three-manifold $M$ is a Seifert
manifold, equipped with a distinguished locally-free $U(1)$ action, and the 
curve $C$ is a Seifert fiber of $M$.  By definition, a locally-free $U(1)$ action is one for which the generating vector field is nowhere vanishing.  Equivalently, all stabilizers under a locally-free $U(1)$ action are proper, necessarily discrete, subgroups of $U(1)$.

Let me introduce a bit of terminology.  The existence of a locally-free $U(1)$ action implies that the Seifert manifold $M$ decomposes geometrically as the total space of a nontrivial circle bundle over a Riemann surface $\Sigma$,
\begin{equation}\label{SFRT}
\begin{matrix}
&S^1 \longrightarrow M\cr
&\mskip 65mu\big\downarrow\lower 0.5ex\hbox{$^\pi$}\cr
&\mskip 55mu\Sigma\cr
\end{matrix}\,.
\end{equation}
Here $\Sigma$ is allowed to have orbifold points, and the circle
bundle is allowed to be a corresponding orbifold bundle, so long as
$M$ itself is smooth.  

Each Seifert manifold carries a distinguished set of Wilson loop operators which respect the $U(1)$ action by rotations in the fiber of \eqref{SFRT}.
For these Wilson loops, the curve $C$ is an orbit of the given $U(1)$ action, and 
$C$ appears geometrically in $M$ as the $S^1$ fiber over a basepoint 
${\sigma\in\Sigma}$.  Assuming that $\sigma$ is a 
smooth (non-orbifold) point of $\Sigma$, the topology of $C$ in $M$
does not depend upon the choice of $\sigma$, so such Wilson loops are 
determined entirely by the choice of the representation $R$.
Henceforth, we refer to these special Wilson loop operators which wrap
the generic Seifert fibers of $M$ as ``Seifert loop operators'' to
distinguish them from general Wilson loops in $M$, about which we
will also have some things to say.  

As a concrete example, $S^3$ admits countably-many locally-free $U(1)$ actions, each associated to a distinct Seifert presentation.  According to a basic result of Moser \cite{Moser:1971}, the corresponding knots which can be realized as Seifert fibers in $S^3$ are precisely the torus knots.  So the Seifert loop operators in $S^3$ are just the Wilson loop operators which wrap torus knots.

\medskip\ni{\it Some Experimental Evidence}\nopagebreak\medskip

The exactness of the stationary-phase approximation to the Seifert loop path integral was discovered by Lawrence and Rozansky \cite{Lawrence:1999} on an empirical basis, through a detailed analysis of the explicitly-known formulae for $Z(k; C, R)$ in the case ${G = SU(2)}$.\footnote{See \cite{Hansen:2002,Marino:2004} for the generalization to other gauge groups $G$.}  Specifically, after a rather involved series of algebraic manipulations,  Lawrence and Rozansky were able to rewrite $Z(k; C, R)$ very compactly as a finite sum of analytic expressions, each being either a contour integral or the residue of a meromorphic function.  These summands in $Z(k; C, R)$ could then be associated in a one-to-one fashion with the connected components in the moduli space of flat connections on $M$.  Since the flat connections on $M$ are the critical points of the Chern-Simons action, such a form for $Z(k;C,R)$ strongly suggests that the stationary-phase approximation to the Seifert loop path integral is exact.

By way of illustration, Lawrence and Rozansky would rewrite the exact formula for the unknot observable $Z\big(k; \bigcirc, \mbox{\boldmath$j$}\big)$ in \eqref{UNKNOTW} as 
\begin{equation}\label{UNKNOTWII}
\begin{split}
&Z\big(k; \bigcirc, {\mbox{\boldmath$j$}}\big) \,=\\ 
&\qquad \frac{1}{2 \pi i} \e{-\frac{i \pi (1 + j^2)}{2(k+2)}} \int_{-\infty}^{+\infty} \!\! dx \,\, \ch_{\mbox{\footnotesize\boldmath$j$}}\!\biggr(\!\e{\frac{i \pi}{4}}\,
\frac{x}{2}\biggr) \sinh^2{\!\left(\e{\frac{i \pi}{4}} \,\frac{x}{2}\right)}
\exp{\!\left(-{\frac{(k+2)}{8 \pi}} \, x^2\!\right)}\,.
\end{split}
\end{equation}
Here $\ch_{\mbox{\footnotesize\boldmath$j$}}$ is the character of $SU(2)$
associated to the representation ${\mbox{\boldmath$j$}}$,
\begin{equation}
\ch_{\mbox{\footnotesize\boldmath$j$}}(y) \,=\, \frac{\sinh(j \,
  y)}{\sinh(y)} \,=\, \e{(j-1) y} \,+\, \e{(j-3) y} \,+\, \cdots \,+\,
\e{-(j-3) y} \,+\, \e{-(j-1) y}\,,
\end{equation}
and the equality between the expressions in \eqref{UNKNOTW} and
\eqref{UNKNOTWII} follows by evaluating \eqref{UNKNOTWII} as a sum of
elementary Gaussian integrals.

Of course, the only flat connection on $S^3$ is the trivial
connection.  As a result, the aforementioned sum in $Z\big(k; \bigcirc, {\mbox{\boldmath$j$}})$ contains only a single term, given by the integral over the real
variable $x$ in \eqref{UNKNOTWII}.  According to Lawrence and Rozansky, this integral is to be interpreted as the stationary-phase contribution from the trivial connection to the full Wilson loop path integral in \eqref{ZkCR}.

One of the main results in \cite{Beasley:2005,Beasley:2009} is to make the semi-classical interpretation of formulae such as \eqref{UNKNOTWII} completely precise.
Briefly, the contour integral over $x$ arises geometrically as an
integral over the Cartan subalgebra of $SU(2)$, regarded as the group
of constant gauge transformations on $S^3$.  The constant gauge
transformations are the stabilizer of the trivial connection in the
group $\CG$ of all gauge transformations, and the presence of this
stabilizer group plays an important role in the semi-classical
analysis of the Wilson loop path integral.  Moreover, all dependence
on the $SU(2)$ representation ${\mbox{\boldmath$j$}}$ enters the integrand of
\eqref{UNKNOTWII} through the character
$\ch_{\mbox{\footnotesize\boldmath$j$}}$.  As a result, the character $\ch_{\mbox{\footnotesize\boldmath$j$}}$ can be naturally interpreted the avatar of the unknot Wilson loop operator itself when the path integral in \eqref{ZkCR} is reduced
to the contour integral in \eqref{UNKNOTWII}.

As an aside, let me mention two interesting generalizations of the semi-classical formula for $Z\big(k; \bigcirc, {\mbox{\boldmath$j$}}\big)$ in \eqref{UNKNOTWII}.  First, this formula extends directly to gauge groups $G$ other than $SU(2)$, in which case the unknot Wilson loop operator for any irreducible representation $R$ of $G$ reduces naturally to the corresponding character $\ch_R$.
Moreover, the semi-classical formula in \eqref{UNKNOTWII} generalizes in a surprisingly simple way to the arbitrary ${({\bf p},{\bf q})}$-torus knot 
$K_{{\bf p},{\bf q}}$ in $S^3$,
\begin{equation}\label{TORUS}
\begin{split}
&Z\big(k;K_{{\bf p},{\bf q}},\mbox{\boldmath$j$}\big) =\, \frac{1}{2
  \pi i} \frac{1}{\sqrt{{\bf p}{\bf q}}} \, \exp{\!\left[-{\frac{i 
\pi}{2 (k+2)}}\left({\frac{\bf p}{\bf q}} +
{\frac{\bf q}{\bf p}} + {\bf p} {\bf q} \,
(j^2-1)\right)\right]} \,\times\,\\
&\times \, \int_{-\infty}^{+\infty} \! dx \;
\ch_{\mbox{\footnotesize\boldmath$j$}}\!\left(\e{\frac{i\pi}{4}}{\frac{x}{2}}\right)
\sinh{\!\left(\e{\frac{i\pi}{4}}\frac{x}{2 {\bf p}}\right)}
\sinh{\!\left(\e{\frac{i\pi}{4}}\frac{x}{2 {\bf q}}\right)}
\exp{\!\left[-{\frac{(k+2)}{8\pi}}
\left(\frac{x^2}{{\bf p}{\bf q}}\right)\right]}.
\end{split} 
\end{equation}
Modulo a rather subtle overall phase, the integrand in \eqref{TORUS} merely acquires denominators proportional to ${\bf p}$ and ${\bf q}$.  From the semi-classical perspective, these denominators are associated to the $U(1)$ stabilizers of the exceptional Seifert fibers in $S^3$.  I refer the interested reader to \S $7.2$ in \cite{Beasley:2009} for a detailed discussion of \eqref{TORUS}.

\medskip\ni{\it Non-Abelian Localization}\nopagebreak\medskip

My goal here is to explain a second, more conceptual way to understand the
exactness of the stationary-phase approximation to the Seifert loop
path integral.  In this approach, we apply non-abelian localization, as
introduced by Witten in \cite{Witten:1992}, to study Chern-Simons
theory on a Seifert manifold.

Very briefly, non-abelian localization provides a cohomological
interpretation for a special class of symplectic integrals which are
intimately related to symmetries.  These integrals take the canonical form 
\begin{equation}\label{PZSM}
Z(\epsilon) \,= \int_X \!
\exp{\!\left[\Omega - \frac{1}{2 \epsilon} \left(\mu,\mu\right)
\right]}.
\end{equation}
Here $X$ is an arbitrary symplectic manifold with symplectic
form $\Omega$.  We assume that a Lie group $H$ acts on $X$ in a
Hamiltonian fashion with moment map ${\mu: X \to \Fh^*}$, where $\Fh^*$
is the dual of the Lie algebra $\Fh$ of $H$.  We also introduce an 
invariant quadratic form $(\,\cdot\,,\,\cdot\,)$ on $\Fh$ and dually on
$\Fh^*$ to define the function ${S = \ha (\mu,\mu)}$ appearing in the
integrand of $Z(\epsilon)$.  Finally, ${\epsilon\in\BR}$ is a coupling
parameter.\footnote{To make sense of the measure on $X$ in \eqref{PZSM}, we expand the exponential $\exp{\!(\Omega)}$ in its Taylor series and pick out
the term $\frac{1}{n!} \, \Omega^n$ of proper degree to integrate over
$X$.  Hence $\exp{\!(\Omega)}$ conveniently describes the symplectic
measure on $X$.}

The symplectic integral in \eqref{PZSM} has a number of important
properties, which for sake of brevity I merely state.  See
\cite{Paradan:2000,Witten:1992} for proofs of the following
statements.  First, the integrand of \eqref{PZSM} admits an
interpretation in terms of the $H$-equivariant cohomology ring of $X$.
Using this interpretation, one can then show that the symplectic
integral itself localizes onto the critical points of the invariant
function ${S = \ha (\mu,\mu)}$ on $X$.  Moreover, a non-abelian
localization formula, roughly analogous to the Duistermaat-Heckman
formula \cite{Duistermaat:1983}, exists to describe the contributions
from the critical locus of $S$.  In a smooth situation, these
contributions are given by the integrals of certain de Rham cohomology
classes over the critical loci.  See \S $6$ of \cite{Beasley:2009}
for a precise statement of the non-abelian localization formula applicable 
to Chern-Simons theory.

\medskip\ni{\it Example: Two-Dimensional Yang-Mills Theory}\nopagebreak\medskip

Given the special form of $Z(\epsilon)$, one should not be surprised
that this integral has special properties.  But why consider such an
integral in the first place?  One answer, following Witten \cite{Witten:1992}, is that the path integral of two-dimensional Yang-Mills theory assumes precisely the canonical symplectic form in \eqref{PZSM}.

To explain the latter observation, I will simply exhibit the
counterparts of $X$, $\Omega$, $H$, and $\mu$ relevant to describe
Yang-Mills theory on a Riemann surface $\Sigma$.  The Yang-Mills path
integral is formally an integral over the affine space $\CA$ of
connections on a fixed principal $G$-bundle $P$ over $\Sigma$, so
clearly we must set 
\begin{equation}
X\,=\,\CA\,.
\end{equation}
The affine space $\CA$ also possesses a natural symplectic form
$\Omega$ given by the intersection pairing on $\Sigma$,
\begin{equation}
\Omega\,=\, -\int_\Sigma \Tr\big(\delta A \^ \delta A\big)\,.
\end{equation}
Here $\delta$ denotes the exterior derivative acting on $\CA$.  Since
$A$ serves as a coordinate on $\CA$, $\delta A$ is a one-form on
$\CA$, and $\Omega$ is a two-form on $\CA$ which is manifestly
non-degenerate and closed.  Of course on $\Sigma$ itself, $\delta A$
transforms as a section of the bundle $\Omega^1_\Sigma \otimes \ad(P)$
of adjoint-valued one-forms.

The obvious group which acts on $\CA$ is the group $\CG$ of gauge
transformations.  As shown long ago by Atiyah and Bott
\cite{Atiyah:1982}, the action of $\CG$ on $\CA$ is Hamiltonian with 
moment map given by the curvature ${F_A = dA + A \^ A}$.  That
is, since elements in the Lie  algebra of $\CG$ appear on $\Sigma$ as
sections of the adjoint bundle $\ad(P)$, the curvature $F_A$, as a
section of $\Omega^2_\Sigma \otimes \ad(P)$, can naturally be
considered as a function on $\CA$ taking values in the dual of the Lie
algebra of $\CG$.  Thus, 
\begin{equation}
H\,=\,\CG\,,\qquad\qquad \mu\,=\,F_A\,.
\end{equation}

Finally, the Lie algebra of $\CG$ admits an invariant form given by 
\begin{equation}\label{LIEG}
(\phi,\phi) \,=\, -\int_\Sigma \Tr\big(\phi \^ \*\phi\big)\,.
\end{equation}
Here $\phi$ is an element of the Lie algebra of $\CG$, transforming on
$\Sigma$ as a section of $\ad(P)$.  With the quadratic form in \eqref{LIEG}, the invariant function ${S =  \ha(\mu,\mu)}$ appearing in the canonical
symplectic integral over $\CA$ immediately becomes the Yang-Mills action,
\begin{equation}\label{YMS}
S \,=\, \ha (\mu,\mu) \,=\, - \ha \int_\Sigma \! \Tr\big(F_A \^ \*
F_A\big)\,.
\end{equation}

The metric on $\Lie(\CG)$ in \eqref{LIEG} is defined using a duality
operator $\*$ on $\Sigma$.  For two-dimensional Yang-Mills theory,
$\*$ relates zero-forms to two-forms, and to obtain such an operator,
we only require a symplectic structure, as opposed to a metric, on
$\Sigma$.  Given a symplectic form $\omega$ on $\Sigma$, we define 
$\*$ by the condition ${\* 1 = \omega}$.  The symplectic form $\omega$
is invariant under all area-preserving diffeomorphisms of $\Sigma$,
and this large group acts as a symmetry of two-dimensional Yang-Mills
theory.  As a result, two-dimensional Yang-Mills theory is essentially
a topological gauge theory.

In the remainder of this article, I want to explain how to recast the
Seifert loop path integral \eqref{ZkCR} as a symplectic integral of
the canonical form \eqref{PZSM}.  Once this step is accomplished, the
general arguments in \cite{Witten:1992} imply that the Seifert loop
path integral localizes onto critical points of the classical action
${S=\ha(\mu,\mu)}$.  Furthermore, using the non-abelian localization
formula, one can perform exact computations of the Seifert loop path
integral and thus obtain a cohomological description for the Seifert
loop operator itself.  Specifically, as demonstrated in \S $7.3$ of 
\cite{Beasley:2009}, the Seifert loop operator reduces naturally to the Chern character of an associated universal bundle over the moduli space of flat connections on $M$.

Two key ideas are required to obtain a symplectic description of the
Seifert loop path integral.  The first idea, which appears in \S $3$
of \cite{Beasley:2005}, pertains to the basic Chern-Simons path
integral in \eqref{CSZ} and really has nothing to do with the Wilson loop
operator.  In contrast, the second idea concerns the Wilson loop
operator itself and really has nothing to do with Chern-Simons theory.
Nonetheless, both of these ideas fit together in a very elegant way.

\section{The Symplectic Geometry of Chern-Simons Theory}

The path integral which describes the partition function of
two-dimensional Yang-Mills theory automatically assumes the canonical
symplectic form in \eqref{PZSM}.  As a special case of our general
study of $Z(k;C,R)$, I now want to review how the path integral
\eqref{CSZ} which describes the partition function $Z(k)$ of
Chern-Simons theory on a Seifert manifold $M$ can also be cast as such 
a symplectic integral.

In order to obtain a symplectic interpretation of the two-dimensional
Yang-Mills path integral, we found it necessary to introduce a
symplectic structure on $\Sigma$.  To obtain a corresponding symplectic
interpretation for the Chern-Simons path integral, we must introduce
the analogous geometric structure on the three-manifold $M$ -- namely,
a contact structure.  

Globally, a contact structure on $M$ is described by an ordinary
one-form $\kappa$, a section of $\Omega^1_M$, which at each point of
$M$ satisfies the contact condition 
\begin{equation}\label{CNTC}
\kappa \^ d\kappa \,\neq\, 0\,.
\end{equation}
By a classic theorem of Martinet \cite{Martinet:1971}, any compact,
orientable\footnote{Any three-manifold admitting a contact structure
  must be orientable, since the nowhere vanishing three-form
  $\kappa\^d\kappa$ defines an orientation.} three-manifold admits a
contact structure, so we do not necessarily assume $M$ to be Seifert
at this stage.  However, if $M$ is a Seifert manifold, then we
certainly want the contact form $\kappa$ to respect the $U(1)$ action
on $M$.  Such a contact form can be exhibited as follows.

We recall that the Seifert manifold $M$ is the total space of an
$S^1$-bundle of degree $n$ over $\Sigma$, 
\begin{equation}\label{SFRTII}
\begin{matrix}
&S^1\,\buildrel n\over\longrightarrow\,M\cr
&\mskip 70mu\big\downarrow\lower 0.5ex\hbox{$^\pi$}\cr
&\mskip 60mu\Sigma\cr
\end{matrix}\,,
\end{equation}
or an orbifold version thereof.  For simplicity, I will phrase the
following construction of $\kappa$ in the language of smooth manifolds,
but the orbifold generalization is immediate.  

Regarding $M$ as the total space of a principal $U(1)$-bundle, we take
$\kappa$ to be a $U(1)$-connection on this bundle which satisfies 
\begin{equation}\label{EULR}
d\kappa\,=\, n\,\pi^*(\omega)\,.
\end{equation}
Here $\omega$ is any unit-area symplectic form on $\Sigma$, and we
recall that a $U(1)$-connection on $\Sigma$ appears upstairs on $M$ as
an ordinary one-form.  Because $d\kappa$ represents the Euler
class of the $S^1$-bundle over $\Sigma$, the degree $n$ necessarily
appears in \eqref{EULR}.  As an abelian connection, $\kappa$ is
automatically invariant under the $U(1)$-action on $M$.  Also, since
the pullback of $\kappa$ to each $S^1$ fiber is non-vanishing, the
contact condition in \eqref{CNTC} is satisfied so long as ${n\neq 0}$
and the bundle is non-trivial, as we assume.

Chern-Simons theory is often considered to be an 
intrinsically three-dimensional gauge theory.  However, one of the
more interesting results in \cite{Beasley:2005, Beasley:2009}
is to show that Chern-Simons theory is {\em not quite} a
three-dimensional gauge theory, since one of the three components of
$A$ can be completely decoupled from all topological observables.

In order to decouple one component of $A$ from the Chern-Simons path
integral, we introduce a new, infinite-dimensional ``shift'' symmetry
$\CS$ which acts on $A$ as 
\begin{equation}\label{SHFT}
\delta A \,=\, \sigma\,\kappa\,.
\end{equation}
Here $\sigma$ is an arbitrary adjoint-valued scalar, a section of
$\Omega^1_M \otimes \Fg$, that parametrizes the action of $\CS$ on
$\CA$.

Of course, the Chern-Simons action $\RC\RS(\,\cdot\,)$ does not
respect the shift of $A$ in \eqref{SHFT}, so we must play a little
path integral trick, of the sort familiar from path integral
derivations of $T$-duality or abelian $S$-duality.  See \S $8$ in
\cite{Deligne:1999} for a nice review of the path integral derivations
of these dualities.

We first introduce a new field $\Phi$ which transforms like $\sigma$
as an adjoint-valued scalar, a section of ${\Omega^1_M\otimes\Fg}$, and
which is completely gauge-trivial under $\CS$.  Thus $\CS$ acts on
$\Phi$ as  
\begin{equation}\label{SPHI} 
\delta \Phi \,=\, \sigma\,.
\end{equation}

Next, we consider a new, shift-invariant action $S(A,\Phi)$
incorporating both $A$ and $\Phi$ such that, if $\Phi$ is set
identically to zero via \eqref{SPHI}, then $S(A,\Phi)$ reduces to the 
Chern-Simons action for $A$.  This condition fixes $S(A,\Phi)$ to be 
\begin{align}\label{SAPI}
S(A,\Phi) \,&=\, \RC\RS(A - \kappa \, \Phi)\,,\cr
&=\, \RC\RS(A) - \int_M \Big[ 2 \kappa \^ \Tr(\Phi
F_A) - \kappa \^ d \kappa \, \Tr(\Phi^2)\Big]\,.
\end{align}

Finally, using \eqref{SAPI} we introduce an {\it a priori} new path
integral\footnote{I will not be very careful about the overall 
  normalizations for the path integrals that appear here, but see \S
  $3$ of \cite{Beasley:2005} for a detailed accounting of formal
  normalization factors.} 
over both $A$ and $\Phi$, 
\begin{equation}\label{WTZKII}
\widetilde Z(k) \,=\, \int \! \CD \! A \, \CD \! \Phi \, \exp{\!\left[i
    {k\over{4\pi}}\, S(A,\Phi)\right]}\,.
\end{equation}
On one hand, if we set ${\Phi \equiv 0}$ using the shift-symmetry in
\eqref{SPHI}, $\widetilde Z(k)$ immediately reduces\footnote{We note that the
  Jacobian associated to the gauge-fixing condition ${\Phi \equiv 0}$
  for $\CS$ is trivial.} 
to the usual Chern-Simons partition function $Z(k)$.  Hence, 
\begin{equation}
\widetilde Z(k) \,=\, Z(k)\,.
\end{equation}

On the other hand, due to the elementary fact that ${\kappa \^ \kappa =
  0}$, the field $\Phi$ appears only quadratically in $S(A, \Phi)$.
So we can simply perform the Gaussian integral over $\Phi$ in
\eqref{WTZKII} to obtain a new path integral description of the
Chern-Simons partition function,
\begin{equation}\label{WTZK}
Z(k) \,=\, \int \! \CD \! A \,  \exp{\!\left[i {k\over{4\pi}}\,
    S(A)\right]}\,,
\end{equation}
where 
\begin{equation}\label{NEWS}
S(A) \,=\, \RC\RS(A) \,-\, \int_M {1\over{\kappa\^d\kappa}}
\Tr\!\left[\left(\kappa\^F_A\right)^2\right]\,.
\end{equation}
In performing the Gaussian integral over $\Phi$, we use the contact
condition on $\kappa$ in \eqref{CNTC}, since this condition ensures
that quadratic term for $\Phi$ in \eqref{SAPI} is everywhere
non-degenerate on $M$.  In particular, the inverse  ``$1 / 
\kappa \^ d\kappa$'' appearing in \eqref{NEWS} is defined as follows.
Because $\kappa \^ d\kappa$ is everywhere non-vanishing, we can always
write ${\kappa \^ F_A = \varphi \, \kappa \^ d\kappa}$ for some section
$\varphi$ of $\Omega^0_M \otimes \Fg$.  Thus, we set ${\kappa \^ F_A /
  \kappa \^ d\kappa = \varphi}$, and the second term in $S(A)$ becomes
$\int_M \kappa \^ \Tr\!\left(F_A \varphi\right)$.

By construction, $S(A)$ is invariant under the shift of $A$ in
\eqref{SHFT}, as can be checked directly.  Thus, we have obtained
a new description of the Chern-Simons partition function for which one
component of $A$ completely decouples from the path integral.
Further, we have yet to use the condition that $M$ be a Seifert
manifold, so the reformulation of $Z(k)$ via the shift-invariant
action in \eqref{NEWS} holds for any three-manifold $M$ endowed with a
contact structure.  See \cite{Jeffrey:2010} for a detailed analysis of the shift-invariant reformulation of $Z(k)$ in the special case that the gauge group is $U(1)$.

\medskip\ni{\it Symplectic Data}\nopagebreak\medskip

If $M$ is a Seifert manifold, an additional miracle occurs, and the
path integral in \eqref{WTZK} becomes an integral of the canonical
symplectic form to which non-abelian localization applies.  For sake
of time, let me merely summarize the symplectic data associated to
\eqref{WTZK}.

First, the space over which we integrate in \eqref{WTZK} and which
must play the role of $X$ is the quotient of the affine space $\CA$ by
the group $\CS$,
\begin{equation}\label{QUOTA}
X \,=\, \CA / \CS\,.
\end{equation}
In three dimensions, the affine space $\CA$ is not symplectic.
However, once we take the quotient by $\CS$ in \eqref{QUOTA}, the
space $\CA/\CS$ carries a natural symplectic form $\Omega$ given by 
\begin{equation}
\Omega \,=\, -\int_M \kappa \^ {\Tr}\!\left(\delta A \^ \delta
  A\right)\,.
\end{equation}
Clearly $\Omega$ is both gauge-invariant and shift-invariant, and
$\Omega$ descends to a non-degenerate symplectic form on $\CA/\CS$.

We must now find a Hamiltonian group acting on $\CA/\CS$ such that
the shift-invariant action $S(A)$ is the square of the corresponding 
moment map.  As an initial guess, motivated by the example of 
two-dimensional Yang-Mills theory, one might consider the group
$\CG$ of gauge transformations.  However, this guess cannot be
correct.  By construction, the square of the moment map for the action
of $\CG$ on $\CA/\CS$ would be invariant under $\CG$.  However, $S(A)$
is plainly not invariant under $\CG$, since the Chern-Simons term 
appearing in \eqref{NEWS} is not invariant under the large gauge
transformations in $\CG$ (while the remaining term in \eqref{NEWS}
is manifestly invariant under $\CG$).

As a second guess, one might replace $\CG$ with its identity component
$\CG_0$, which does preserve the shift-invariant action $S(A)$.
However, one can show that $\CG_0$ is obstructed from acting in a
Hamiltonian fashion on $\CA/\CS$ by a non-trivial Lie algebra cocycle
$c$, 
\begin{equation}\label{COC}
c(\phi,\psi) \,=\, -\int_M d\kappa \^ \Tr\!\left(\phi \, d\psi\right)\,.
\end{equation}
Here $\phi$ and $\psi$ are elements of the Lie algebra of $\CG_0$,
transforming as sections of $\Omega^0_M\otimes\Fg$ on $M$.
Parenthetically, this cocycle is closely related to a cocycle that
appears in the theory of loop groups \cite{Pressley:1986}, which
also provides useful background for the identification of $H$ below.

To remedy the situation, we consider the central extension\footnote{A Lie
  algebra two-cocycle always determines a central extension of algebras.
  Provided that the cocycle is properly quantized, as is the cocycle
  in \eqref{COC}, the central extension of algebras lifts to a
  corresponding central extension of groups.}
$\widetilde\CG_0$ of $\CG_0$ determined by $c(\phi,\psi)$,
\begin{equation}\label{ZNTRLG}
 U(1)_\RZ \buildrel c\over\longrightarrow {\widetilde\CG}_0 \longrightarrow
\CG_0\,.
\end{equation}
Here we use the subscript `$\RZ$' to emphasize that $U(1)_\RZ$ is
central in ${\widetilde\CG}_0$.  The natural action of $\CG_0$ on $\CA/\CS$
extends to an action by ${\widetilde\CG}_0$, for which $U(1)_\RZ$ acts
trivially.  By construction, the action of ${\widetilde\CG}_0$ on $\CA/\CS$
is then Hamiltonian.

However, ${\widetilde\CG}_0$ is still not the group which is to play the role
of the Hamiltonian group $H$!  In order to define the canonical
symplectic integral in \eqref{PZSM}, the Lie algebra of $H$ must carry
a non-degenerate, invariant quadratic form $(\,\cdot\,,\,\cdot\,)$.
But the Lie algebra of ${\widetilde\CG}_0$ does not admit such a form,
essentially because we have no generator to pair with the generator of
the central $U(1)_\RZ$.  

However, we have also yet to apply the Seifert condition on $M$.  We
do so now.  To avoid confusion, let me denote the Seifert 
$U(1)$ acting on $M$ by $U(1)_\RR$, to distinguish it from the central
$U(1)_\RZ$.  The action by $U(1)_\RR$ on $M$ induces a corresponding
action on both ${\widetilde\CG}_0$ and $\CA/\CS$, and we finally take $H$ to
be the semi-direct product 
\begin{equation}\label{BIGH}
H \,=\, U(1)_\RR \ltimes {\widetilde\CG}_0\,.
\end{equation}

The Lie algebra of $H$ does admit a non-degenerate, invariant
quadratic form $(\,\cdot\,,\,\cdot\,)$, under which the generators of
$U(1)_\RR$ and $U(1)_\RZ$ are paired.  Furthermore, the action of $H$
on $\CA/\CS$ is Hamiltonian with moment map $\mu$ (for which a
completely explicit though perhaps not so illuminating formula
exists), and the corresponding invariant function ${S=\ha(\mu,\mu)}$
on $\CA/\CS$ is precisely $S(A)$.  Given the amount of symmetry
respected by both $\ha(\mu,\mu)$ and $S(A)$, the latter result could
hardly have been otherwise.

The presentation here is a regrettably quick sketch of a fairly
miraculous result, and I refer the reader to \S $3$ of
\cite{Beasley:2005} for a complete discussion.

\section{Inclusion of Wilson Loop Operators}

I now want to explain how the prior statements concerning the
Chern-Simons partition function can be generalized to allow for
insertions of Wilson loop operators.  (See \S $4$ of \cite{Beasley:2009} for an expanded version of the material here.)  As it happens, only one new idea 
is required. 

We clearly need a new idea, because a naive attempt to reapply the
previous path integral manipulations to the Wilson loop path integral in
\eqref{ZkCR} runs immediately aground.   To illustrate the
difficulty with the direct approach, let us consider the obvious way
to rewrite the Wilson loop path integral in a shift-invariant form,
\begin{equation}\label{PZCSWLII}
Z(k; C, R) \,=\, \int \CD \! A \, \CD \! \Phi
\,\, \CW_{\!R}(C) \,\exp{\left[i {k\over{4\pi}} \RC\RS(A - \kappa \,
    \Phi)\right]}\,.
\end{equation}
Here $\CW_{\!R}(C)$ denotes the generalized Wilson loop operator defined
not using $A$ but using the shift-invariant combination $A - \kappa
\, \Phi$, so that
\begin{equation}\label{SHFTWL} 
\CW_{\!R}(C) \,=\, {\Tr}_R \, P\exp{\!\left[-\oint_C \!\big(A -
\kappa \, \Phi\big)\right]}\,.
\end{equation}
Exactly as for our discussion of \eqref{WTZKII}, we can use the shift
symmetry to fix ${\Phi \equiv 0}$, after which the path integral in
\eqref{PZCSWLII} reduces trivially to the standard Wilson loop path
integral in \eqref{ZkCR}.

However, to learn something useful from \eqref{PZCSWLII} we must
perform the path integral over $\Phi$, and as it stands, this integral
is not easy to do.  Because the generalized Wilson loop operator 
$\CW_{\!R}(C)$ is expressed in \eqref{SHFTWL} as a complicated,
non-local functional of $\Phi$, the path integral over $\Phi$ in
\eqref{PZCSWLII} is not a Gaussian integral that we can trivially
evaluate as we did for \eqref{WTZKII}.

A more fundamental perspective on our problem is the following.  Let
us return to the description of the ordinary Wilson loop operator  
$W_R(C)$ as the trace in the representation $R$ of the holonomy of
$A$ around $C$,
\begin{equation}\label{DEFWVCII}
W_R(C) \,=\, {\Tr}_R \, P\exp{\left(-\oint_C\! A\right)}\,.
\end{equation}
As observed by Witten in one of the small gems of \cite{Witten:1988},
this description of $W_R(C)$ should be regarded as intrinsically
quantum mechanical, for the simple reason that $W_R(C)$ can be
naturally interpreted in \eqref{DEFWVCII} as the partition function of
an auxiliary quantum system attached to the curve $C$.  Briefly, the
representation $R$ is to be identified with the Hilbert space of this
system, the holonomy of $A$ is to be identified with the
time-evolution operator around $C$, and the trace over $R$ is the
usual trace over the Hilbert space  that defines the partition
function in the Hamiltonian formalism.

Because the notion of tracing over a Hilbert space is inherently
quantum mechanical, any attempts to perform essentially classical path
integral manipulations involving the expressions in \eqref{SHFTWL} or
\eqref{DEFWVCII} are misguided at best.  Rather, if we hope to generalize 
the semi-classical path integral manipulations which we used
to study the Chern-Simons partition function, we need to use an 
alternative, semi-classical description for the Wilson loop operator itself.

More precisely, we want to replace the quantum mechanical trace over
$R$ in \eqref{DEFWVCII} by a path integral over an auxiliary bosonic
field $U$ which is attached to the curve $C$ and coupled to the
connection $A$ as a background field, so that schematically 
\begin{equation}\label{NEWWC}
 W_R(C) \,=\, \int \!\CD\!U \, \exp{\Big[
i\,\Rc\Rs_\alpha\!\big(U; A|_C\big)\Big]}\,.
\end{equation}
Here $\Rc\Rs_\alpha\!\big(U; A|_C\big)$ is an action, depending upon the
representation $R$ through its highest weight $\alpha$, which is a local, 
gauge-invariant, and indeed topological functional of the defect field $U$ 
and the restriction of $A$ to $C$.  Not surprisingly, this semi-classical 
description \eqref{NEWWC} of $W_R(C)$ turns out to be the key ingredient 
required to reformulate the Wilson loop path integral in a shift-invariant
fashion.

The idea of representing the Wilson loop operator by a path integral
as in \eqref{NEWWC} is a very old and very general piece of gauge
theory lore.  In the context of four-dimensional Yang-Mills theory,
this idea goes back (at least) to work of Balachandran, Borchardt, and
Stern \cite{Balachandran:1978} in the 1970's.  See also
\cite{Diakonov:1989} and \S $7.7$ in \cite{Deligne:1999} for other
appearances of the path integral in \eqref{NEWWC}.

The basic idea behind the path integral description \eqref{NEWWC} of the Wilson loop operator is very simple.  We interpret the closed curve $C$ as a periodic ``time'' for the field $U$, and we apply the Hamiltonian formalism to rewrite the path
integral over $U$ axiomatically as the quantum mechanical trace of the
corresponding time-evolution operator around $C$,
\begin{equation}\label{ZC}
W_R(C) \,=\, {\Tr}_\SH \, P\exp{\left(-i \oint_C\!{\bf H}\right)}\,.
\end{equation}
Here $\SH$ is the Hilbert space obtained by quantizing $U$, and
${\bf H}$ is the Hamiltonian which acts on $\SH$ to generate
infinitesimal translations along $C$.

Comparing the conventional description of the Wilson loop operator in
\eqref{DEFWVCII} to the axiomatic expression in \eqref{ZC}, we see
that the two agree if we identify\footnote{We follow the standard physical definition
according to which ${\bf H}$ is a hermitian operator, accounting for
the `$-i$' in \eqref{ZC}.  We also recall that the gauge field $A$ is valued
in the Lie algebra $\Fg$, so $A$ is anti-hermitian and no `$i$'
appears in the holonomy.}
\begin{align}\label{ISQM}
R \,&\longleftrightarrow\,\SH\,,\cr
P\exp{\left(-\oint_C\! A\right)}\,&\longleftrightarrow\, P
\exp{\left(-i \oint_C\!{\bf H}\right)}\,.
\end{align}
Hence to make the Wilson loop path integral in \eqref{NEWWC} precise,
we need only exhibit a classical theory on $C$, for which the gauge group $G$
acts as a symmetry, such that upon quantization we obtain a Hilbert
space $\SH$ isomorphic to $R$ and for which the time-evolution
operator around $C$ is given by the holonomy of $A$, acting as an
element of $G$ on $R$.

\medskip\ni{\it A Semi-Classical Description of the Wilson Loop Operator}\nopagebreak\medskip

Let me now tell you what classical theory to place on $C$ to realize the
quantum identifications in \eqref{ISQM}.

Of the two identifications in \eqref{ISQM}, the more fundamental by far
is the identification of the irreducible representation $R$ with a
Hilbert space $\SH$, obtained by quantizing some classical phase space upon
which $G$ acts as a symmetry.  So before we even consider what
classical theory must live on $C$ to describe the Wilson loop
operator, we can ask the simpler and more basic question ---
what classical phase space must we quantize to obtain $R$ as a
Hilbert space?

As well-known, this question is beautifully answered by the
Borel-Weil-Bott theorem \cite{Bott:1957}.  In order to recall this
theorem, let me first fix a maximal torus ${T \subset G}$, for which
${\Ft \subset \Fg}$ is the associated Cartan subalgebra.  Given the
irreducible representation $R$ and some choice of positive roots for
$G$, I also introduce the associated highest weight $\alpha$.
Canonically, the weight $\alpha$ lies in the dual $\Ft^*$ of $\Ft$,
but given the invariant form `$\Tr$' on $\Fg$, we are free to identify
${\Ft^* \cong \Ft}$ and hence to regard $\alpha$ as an element of $\Ft$,
\begin{equation}\label{CONVT}
\alpha \in \Ft^* \cong \Ft\,.
\end{equation}
Though mathematically unnatural, the convention in \eqref{CONVT} proves to be  convenient later.

The Borel-Weil-Bott theorem concerns the geometry of the orbit
${\CO_\alpha \subset \Fg}$ which passes through $\alpha$ under the
adjoint action of $G$.  Equivalently, the adjoint orbit  $\CO_\alpha$ can 
be realized as a quotient $G/G_\alpha$, where $G_\alpha$ is 
the stabilizer of $\alpha$ under the adjoint action of $G$.  Explicitly, the
identification between $G/G_\alpha$ and $\CO_\alpha$ is given by
the map  
\begin{equation}\label{MOMENT}
g \, G_\alpha \,\longmapsto\, g\,\alpha\,g^{-1}\,,\qquad g\in G\,.
\end{equation}
As will be essential in a moment, $\CO_\alpha$ is a compact complex
manifold which admits a natural K\"ahler structure invariant under
$G$.  For instance, if ${G = SU(2)}$ and $\alpha$ is any
non-zero\footnote{Of course, if $\alpha = 0$, then $\CO_\alpha$ is
  merely a point.} weight, then ${\CO_\alpha = SU(2)/U(1)}$ can be identified
as $\BC\BP^1$ endowed with the round, Fubini-Study metric.

In a nutshell, the Borel-Weil-Bott theorem states that the irreducible
representation $R$ can be realized geometrically as the space of
holomorphic sections of a certain unitary line bundle $\FL(\alpha)$
over $\CO_\alpha$.  That is, 
\begin{equation}\label{BWBT}
R \,\cong\, H^0_{\bar\partial}\big(\CO_\alpha,\,\FL(\alpha)\big)\,,
\end{equation} 
where the action of $G$ on the sections of $\FL(\alpha)$ is induced from
its action on $\CO_\alpha$. 

As a unitary line bundle over a K\"ahler manifold, $\FL(\alpha)$
carries a natural unitary connection $\Theta_\alpha$ which is also
invariant under $G$.  The connection $\Theta_\alpha$ enters the path
integral description of the Wilson loop operator, so let me exhibit it
explicitly.  When pulled back to $G$, the line bundle $\FL(\alpha)$
trivializes, and the connection $\Theta_\alpha$ appears as the
following left-invariant one-form on $G$,
\begin{equation}\label{BIGTH}
\Theta_\alpha \,=\, \Tr\big(\alpha \cdot g^{-1} dg\big)\,.
\end{equation}
I have introduced the connection $\Theta_\alpha$ because its curvature
${\nu_\alpha = d\Theta_\alpha}$ is precisely the K\"ahler form on
$\CO_\alpha$.  As a result, $\FL(\alpha)$ is a prequantum line bundle
over $\CO_\alpha$, and the Borel-Weil-Bott isomorphism in \eqref{BWBT}
identifies $R$ as the Hilbert space obtained by K\"ahler quantization
of $\CO_\alpha$.

Perhaps more physically, the Borel-Weil-Bott theorem can be
interpreted as identifying the space of groundstates for a
charged particle moving on $\CO_\alpha$ in the presence of a
background magnetic field given by $\nu_\alpha$.  Briefly, because of the
non-zero magnetic field, the wavefunctions which describe this
particle transform on $\CO_\alpha$ not as functions but as sections of
the line bundle $\FL(\alpha)$.  As standard, the Hamiltonian which
describes free propagation on $\CO_\alpha$ is proportional to the
Laplacian $\triangle$ acting on sections of $\FL(\alpha)$, and by
Hodge theory, the kernel of $\triangle$ can be identified as the space
of holomorphic sections of $\FL(\alpha)$.  Hence the role of
\eqref{BWBT} is to realize the representation $R$ in terms of groundstates on $\CO_\alpha$.

Given the previous quantum mechanical interpretation for $R$, the
corresponding path integral description \eqref{NEWWC} for the Wilson
loop operator follows immediately.  Ignoring the coupling to $A$ for a
moment, if we simply wish to describe the low-energy effective
dynamics of an electron moving on $\CO_\alpha$ in the background
magnetic field $\nu_\alpha$, we consider a one-dimensional sigma
model of maps 
\begin{equation}\label{BIGU}
U:C \longrightarrow  \CO_\alpha\,,
\end{equation}
with sigma model action 
\begin{equation}\label{CSRI}
\Rc\Rs_\alpha(U) \,=\, \oint_C U^*(\Theta_\alpha) \,=\, \oint_C
\Tr(\alpha \cdot g^{-1} dg)\,.
\end{equation}
Here $U^*(\Theta_\alpha)$ denotes the pullback of $\Theta_\alpha$ to a
connection over $C$.  If $U$ is lifted as a map to 
${\CO_\alpha = G/G_\alpha}$ by a corresponding map 
\begin{equation}
g:C \longrightarrow G\,,
\end{equation}
then the pullback of $\Theta_\alpha$ appears explicitly as in
\eqref{CSRI}.  As a word of warning, I will freely switch between
writing formulae in terms of $U$ or $g$ as convenient.

From a physical perspective, the first-order action $\Rc\Rs_\alpha$
simply describes the minimal coupling of the charged particle on
$\CO_\alpha$ to the background magnetic field specified by
$\Theta_\alpha$, and we have omitted the second-order kinetic terms
for $U$ as being irrelevant at low energies.  From a more geometric
perspective, $\Rc\Rs_\alpha$ is a one-dimensional Chern-Simons action for
the abelian connection $U^*(\Theta_\alpha)$ over $C$.  As such, the
quantization of the parameter ${\alpha \in \Ft}$ as a weight of $G$
follows just as for the quantization of the Chern-Simons level $k$.
More physically, the quantization of $\alpha$ follows from the quantization
of magnetic flux on a compact space.

This sigma model on $C$ clearly respects the action of $G$ on
$\CO_\alpha$ as a global symmetry.  To couple the sigma model to the 
restriction of the bulk gauge field $A$, we simply promote the global
action of $G$ on $\CO_\alpha$ to a gauge symmetry.  That is, we
consider the covariant version of \eqref{CSRI},
\begin{align}\label{CSRII}
\Rc\Rs_\alpha(U;A|_C) \,&=\, \oint_C U^*(\Theta_\alpha(A)) \,=\, \oint_C
\Tr(\alpha \cdot g^{-1} d_A g)\,,\cr
d_A g \,&=\, dg + A|_C \cdot g\,.
\end{align}
In the second line, we indicate the action of the covariant derivative
$d_A$ on $g$.  The action by $d_A$ on $g$ descends to a corresponding
covariant action by $d_A$ on $U$ as well.

I now claim that the quantization of the gauged sigma model on $C$ with
action $\Rc\Rs_\alpha(U;A|_C)$ leads to the identifications in \eqref{ISQM}
required to describe the Wilson loop operator.  First, the classical
equation of motion for $U$ simply asserts that $U$ is covariantly
constant,
\begin{equation}
d_A U \,=\, 0\,.
\end{equation}
As a result, the classical phase space for $U$ can be identified with
the orbit $\CO_\alpha$, and by the Borel-Weil-Bott isomorphism in 
\eqref{BWBT}, the corresponding Hilbert space $\SH$ for $U$ is
identified as the representation $R$.  Similarly, since the classical
time-evolution for $U$ is given by parallel transport, the quantum
time-evolution operator around $C$ is immediately given by the
holonomy of $A$, acting as an element of $G$ on $R$.

\medskip\ni{\it The Shift-Invariant Wilson Loop in Chern-Simons Theory}\nopagebreak\medskip

Because $A$ only enters as a background field in \eqref{CSRII}, the path
integral description \eqref{NEWWC} of $W_R(C)$ is completely
general and applies to any gauge theory in any dimension.
Nonetheless, this description of the Wilson loop operator is precisely  
what we need to obtain a shift-invariant formulation of the Wilson
loop path integral in Chern-Simons theory.

Let us first apply \eqref{NEWWC} to rewrite the Wilson loop path
integral in \eqref{ZkCR} as a path integral over both $A$ and $U$,
\begin{equation}\label{ZkCRI}
Z(k; C, R) \,=\, {1\over{\Vol(\CG)}} \int_{\CA \times L\CO_\alpha} \!
\CD \! A \, \CD \! U \; \exp{\left[i {k \over {4 \pi}} \RC\RS(A)
    \,+\, i \, \Rc\Rs_\alpha(U;A|_C)\right]}\,.
\end{equation}
Here we introduce the free loopspace $L\CO_\alpha$ of $\CO_\alpha$ to
parametrize configurations of $U$.

Once we introduce the defect field $U$ coupling to $A$ in
\eqref{ZkCRI}, the classical equation of motion for $A$ becomes  
\begin{equation}\label{NEWF}
\CF_A \,\buildrel{\rm def}\over =\, F_A + {{2\pi}\over k} \, U \cdot
\delta_C \,=\, 0\,.
\end{equation}
Here $\delta_C$ is a two-form with delta-function support on $C$ which
represents the Poincar\'e dual of the curve.  Using $\delta_C$, we rewrite
$\Rc\Rs_\alpha(U; A|_C)$ as a bulk integral
over $M$,
\begin{equation}\label{BULKM}
\Rc\Rs_\alpha\big(U; A|_C\big) \,=\, \oint_C \Tr\big(\alpha \cdot g^{-1}
d_A g\big) \,=\, \int_M \delta_C \^ \Tr\big(\alpha \cdot g^{-1} d_A
g\big)\,,
\end{equation}
from which \eqref{NEWF} follows.

As we see from \eqref{NEWF}, in the presence of the operator $W_R(C)$,
classical configurations for $A$ are given by connections which are
flat on the knot complement 
\begin{equation}
M^o = M - C\,,
\end{equation} 
and otherwise have
delta-function curvature along $C$.  The singularity in $A$
along $C$ manifests itself on $M^o$ as a non-trivial monodromy of the
connection around a transverse circle linking $C$.  Though I will not
have time to say more, the moduli space of such flat connections with
monodromies is the space onto which the Seifert loop path integral
localizes.  This space is directly related to the moduli space of
representations of the knot group of $C$ in $G$ and, in suitable
circumstances, fibers over the moduli space of (non-singular)
flat connections on $M$.

At the cost of introducing defect degrees-of-freedom along $C$, we
have managed to describe $W_R(C)$ in terms of a completely local 
--- and indeed linear --- functional of $A$.  Consequently, the same
path integral trick that we used to decouple one component of $A$ from
the Chern-Simons partition function applies immediately to
\eqref{ZkCRI}.  We simply replace $A$ in \eqref{ZkCRI} by the
shift-invariant\footnote{The shift-symmetry $\CS$ acts trivially on
  $U$.} combination ${A - \kappa\,\Phi}$, and we then perform the 
Gaussian integral over $\Phi$. In the process, the only new ingredient
is that we obtain an extra term linear in $\Phi$ from the coupling of
$A$ to $U$.

Without discussing any more details, let me present the resulting
shift-invariant formulation for the Wilson loop path integral,
\begin{equation}\label{ZkCRII}
Z(k; C, R) \,=\, {1\over{\Vol(\CG)}} \int_{\CA/\CS \times L\CO_\alpha} \!
\CD \! A \, \CD \! U \; \exp{\left[i {k \over {4 \pi}} S(A,U)
  \right]}\,,
\end{equation}
where
\begin{equation}\label{SAPII}
S(A, U) \,=\, \RC\RS(A) \,+\, {{4\pi}\over k} \Rc\Rs_\alpha(U;A|_C)
\,-\, \int_M {1 \over {\kappa \^ d\kappa}} \Tr\!\left[(\kappa \^
  \CF_A)^2\right]\,.
\end{equation}
The shift-invariant action $S(A, U)$ appears much as the 
shift-invariant action \eqref{NEWS} for $A$ alone, with the
replacement therein of $F_A$ by the generalized curvature $\CF_A$. 
Thus for an arbitrary Wilson loop (or link) operator in Chern-Simons
theory, the path integral can be rewritten such that one component of
$A$ completely decouples.

Moreover, if $M$ is a Seifert manifold and $C$ is a Seifert fiber of
$M$, the shift-invariant Seifert loop path integral is again an
integral of the canonical symplectic form in \eqref{PZSM}.  The relevant
symplectic space $X$ is just the product 
\begin{equation}
X \,=\, \CA/\CS \times L\CO_\alpha\,,
\end{equation}
where the loopspace $L\CO_\alpha$ carries a natural symplectic (and
indeed K\"ahler) form inherited from the K\"ahler form $\nu_\alpha$ on
$\CO_\alpha$.  The Hamiltonian group $H$ which acts on $X$ is the same group that appears in \eqref{BIGH}.  In fact, the loopspace $L\CO_\alpha$ can 
be interpreted formally as an infinite-dimensional coadjoint orbit of
$H$.  Finally, the square of the moment map for the diagonal action of $H$
on $X$ is precisely the shift-invariant action $S(A, U)$ appearing 
in \eqref{SAPII}.

For a last bit of furious hand-waving, let me remark that the
description of the Seifert loop operator as a character follows quite
naturally from the appearance of the loopspace $L\CO_\alpha$ in
\eqref{ZkCRII}.  In essence, non-abelian localization on $L\CO_\alpha$
is related to index theory on $\CO_\alpha$, and index theory on
$\CO_\alpha$ provides a classic derivation \cite{Atiyah:1968} of
the famous Weyl character formula.  See \S $7.2$ in \cite{Beasley:2009} for a complete discussion.

\section*{Acknowledgements}

I take pleasure in thanking the organizers and participants of the Bonn workshop ``Chern-Simons Gauge Theory:~20 Years After," where this article was presented.  I especially thank Edward Witten, both for our prior collaboration on the subject and for posing the question which sparked this work.

\bibliographystyle{amsalpha}

\end{document}